\documentclass[manuscript,nonacm]{acmart}

\usepackage{xcolor, listings, framed}
\usepackage{xspace}

\definecolor{shadecolor}{gray}{0.95}
\AtBeginDocument{%
  }

\begin{document}

\title{Flowcode: An AI-Powered Programming Environment for Scaffolding Iteration in Creative Computing Education}

% \author{Anonymized}

\author{Tiffany Tseng}
\affiliation{%
  \institution{Barnard College, Columbia University}
  \city{New York}
  \country{United States}}

\author{Liliana Hanem Seoror}
\affiliation{%
  \institution{Columbia University}
  \city{New York}
  \country{United States}}

\author{Jeevika Adda}
\affiliation{%
  \institution{Columbia University}
  \city{New York}
  \country{United States}}

\author{Meitalia Factor}
\affiliation{%
  \institution{Barnard College, Columbia University}
  \city{New York}
  \country{United States}}

\author{Rona Darabi}
\affiliation{%
  \institution{Columbia University}
  \city{New York}
  \country{United States}}

\author{Kiley R Matschke}
\affiliation{%
  \institution{Columbia University}
  \city{New York}
  \country{United States}}

\author{Tiffany Fu}
\affiliation{%
  \institution{Columbia University}
  \city{New York}
  \country{United States}}

\author{Annie Lin}
\affiliation{%
  \institution{Barnard College, Columbia University}
  \city{New York}
  \country{United States}}

\author{Alekhya Maram}
\affiliation{%
  \institution{Barnard College, Columbia University}
  \city{New York}
  \country{United States}}

\author{Arya Sinha}
\affiliation{%
  \institution{Barnard College, Columbia University}
  \city{New York}
  \country{United States}}

\renewcommand{\shortauthors}{Tseng, T., Seoror, L.H., Adda, J., Factor, M., Darabi, R., Matschke, K.R., Fu, T., Lin, A., Maram, A., Sinha, A.}

\begin{abstract}
Building upon found examples is a popular way people learn to code, especially in creative coding communities where sharing projects and remixing are common practices. But effectively doing so requires being able to 1) understand how existing code works, and 2) extend it by writing code that implements your own ideas, practices that can be challenging for new creative coders. We explored how to support these two processes through the design of Flowcode, a creative coding programming environment that integrates a flowchart for visualizing code structure and a chat interface tailored to support learning to code over vibe coding. We share how we iterated on the design of Flowcode over two studies with new creative coders, reflecting on the roles visualization and friction may play in enabling productive AI-use in computing education.
\end{abstract}





\authorsaddresses{Corresponding author: Tiffany Tseng, ttseng@barnard.edu, Barnard College, Columbia University, New York, United States.}

\maketitle

\newcommand{\todo}[1]{{\textcolor{red}{[#1]}\normalfont}}
\newcommand{\rqone}{my research question}
\newcommand{\system}{Flowcode}

\section{Introduction}
\label{sec:intro}

Examples play a fundamental role in how people design, helping creators get inspired, learn from prior work, and build off of to create something new \cite{herring2009getting, kulkarni2013early}. % Examples can be especially important for spurring creativity when utilized early in a design process \cite{kulkarni2013early}. 
With software development and design, the web has greatly facilitated the use of examples given the ability to quickly distribute, inspect, and remix code samples \cite{lee2010designing, brandt2010example, brandt2009two, yang2024considering, subbaraman2023forking}. Prior work in computing education has found positive associations between remixing code and exposure to and use of new programming concepts \cite{dasgupta2016remixing}, aligning with the pedagogical practice of providing worked examples \cite{muldner2022review, brusilovsky2001webex}. Furthermore, real-world professional code can serve as powerful examples because they reflect current practices rather than simplified code created for instructional purposes \cite{dasu2026process}, and existing work has explored how interactive visualizations of professional examples support learners' conceptual understanding \cite{hibschman2019isopleth}. However, effectively remixing found samples depends on one's ability to both \textit{understand} code written by another creator, and \textit{iterate} on found examples to realize one's personal design goals, practices that may be especially challenging to students who are still learning to program or utilize unfamiliar programming frameworks \cite{wang2021novices}.
% "Books, blogs, tutorials, courses, and other manually created resources cover this knowledge gap, but they often focus on canonical examples in a single context, without discussing how to apply these techniques across complex real-world scenarios... Professional websites are a potentially powerful sensemaking context because they are inspectable with browser tools, reflect up-to-date techniques, and do not need to be maintained like tutorials"

Our work examines how to foster code understanding and iteration in the context of \textit{creative coding}, or using code to create visual art. With creative coding, there is a strong ethos of sharing and remixing code as exemplified in online communities like OpenProcessing\footnote{https://openprocessing.org/} and CodePen\footnote{https://codepen.io/}, community showcases \cite{p5js}, and integrated examples built into IDEs like Processing \cite{processing}. Prior work has shown that engaging in creative coding can positively affect learners' intentions and interests in computing, particularly for traditionally underrepresented communities \cite{bares2018gender, wood2016computational, payne2021danceon}. Because of this, making it easier for creative coders to understand and build on top of shared creative coding projects may contribute to encouraging more diversity in computing.

%suggested changing "towards" to "toward" because I think "toward" is preferred in American English?
A potential pathway toward supporting effective remixing is leveraging large language models (LLMs) that are capable of generating explanations and code support across a wide range of programming contexts \cite{chen2021evaluating}. However, popular code assistants like GitHub Copilot or Cursor are primarily designed for efficiency rather than learning, with developers commonly using them to accelerate their process of writing code \cite{barke2023grounded}. This has led to reservations about the role of AI in computer science education \cite{lau2023ban, kasneci2023chatgpt}, along with work to instrument guardrails that prevent LLMs from generating complete code solutions in course contexts \cite{kazemitabaar2023studying, liu2024teaching}. However, there is limited work examining the role of code assistants for learning \textit{creative coding} specifically other than \cite{jonsson2022cracking}, which examined how students used ChatGPT out-of-the-box to implement their designs. We believe there is value in combining strategies for pedagogically-grounded code assistance with the unique needs of creative computing (such as the need to flexibly accommodate both design and programming goals in parallel, and the need for tight feedback loops to visualize changes to programs \cite{resnick2013designing}). This has motivated our design of Flowcode, a programming environment that utilizes AI to generate visualizations and explanations to support learners iterating on found creative coding projects. 

In this paper, we share our own iterative process of building Flowcode, reporting on how its design was refined across two user studies with students new to creative coding: a workshop setting (n=7) and a lab study (n=9) where participants iterated on animations using web programming languages. Through this work, we share (1) the role that visualization can play in positively supporting understanding of found code; %particularly with respect to how design features can be specified \textit{across} HTML, CSS, and JavaScript; 
and (2) our approach to introducing friction into the review and use of generated explanations to reduce vibe coding. This work contributes (1) the design of Flowcode and AI features designed for code understanding and supporting iteration, and (2) insights from two empirical studies showing how these features affected learner perception on the role of AI in creative coding education.

\section{Related Work}
\label{sec:related-work}
\subsection{LLMs for Code Comprehension}
Reading code is a fundamental skill within programming. To maintain codebases, professional developers need to regularly read and comprehend code \cite{ko2006exploratory,craig2018listening, robillard2005effective}, skills that novice programmers often experience difficulty doing \cite{simon2009surely, lister2004multi,woerner2023code}, let alone contributing to code \cite{shah2025needles}. As such, computer science education (CSE) has long studied methods of teaching \cite{busjahn2013use,lewis2023examples,hibschman2019isopleth} and assessing  \cite{izu2020comparing,lehtinen2021students,murphy2012explain,simon2014multiple} code comprehension skills.

The emergence of LLMs has generated significant interest around the impact they may have on CSE due to their code generation capabilities \cite{prather2023robots}. While many IDEs now integrate code assistants that help developers quickly debug and add features to their code (such as GitHub Copilot and research prototypes like \cite{nam2024using} and \cite{yan2024ivie}), they often generate and insert code solutions directly \cite{barke2023grounded}, which may run counter to the goal of helping beginners learn to code and lead to vibe-coding \cite{geng2026exploring}. This risk has been observed beyond individual learners: in group projects for programming courses, students have reported submitting AI-generated code they did not fully understand, leaving teammates to spend more time debugging work they could not verify \cite{kharrufa2026llms}. Therefore, code reading is becoming increasingly relevant as students who use LLMs need to read, interpret, and assess code that is generated \cite{picoral2026code, prather2024widening, qiao2026systematic, zi2025would, dasu2026process}.
Several works have highlighted the importance of implementing guardrails within CSE LLM tools to better scaffold students' code comprehension \cite{qiao2026systematic, denny2024desirable, kazemitabaar2023studying, liffiton2023codehelp, kazemitabaar2024codeaid}, utilizing strategies such as diversified code explanations \cite{patel2026assessing} and interactive teachable agents \cite{jin2024teach}. 

Few studies have explored how LLMs can encourage computational thinking within creative coding contexts, where code is a means of creating interactive visual art \cite{levin2021code}. \citet{jonsson2022cracking} studied how design students utilized GPT-3 Codex for creative coding, sharing a concern that GenAI might demotivate students from learning programming given that students can instruct AI to generate designs directly.
% Their findings align with those found in beginner CS learner environments: many participants encountered issues where the GenAI system did not generate code as they expected; rather, their success of getting an intended code response from the system depended on the user's ability to craft prompts in a way that the LLM is able to understand.
However, they do not propose approaches by which LLMs might instead support learning programming concepts in creative coding.
%Wang et al. \cite{wang2025pinning} compared prompting strategies for creative coding but focused more on how prompting can structure creative reflection rather than learning and understanding.
Educators that teach creative coding are mixed in their perspectives about the role of AI, with fear that AI may make creating less satisfying, while others see potential for it to aid in idea generation \cite{mcnutt2025slowness}. 

We see Flowcode as addressing a missing link between LLM-based pedagogical programming tools and the creative coding context. In creative coding, code serves open-ended, constantly shifting design goals rather than more closed-ended programming exercises that converge on a single solution. Further, the visual output is of primary importance, so helping users relate code to visuals is essential, a practice we aim to support by utilizing LLM-generated visualizations of code structure. 

% While they mention that users sought help from LLMs for debugging purposes and code explanations, the focus of their paper was to understand how prompting can structure creative reflection, rather than optimize for learning and understanding. 
 % We propose that a LLM-based tool like FlowCode could be beneficial in helping novices learn about the behavior of unfamiliar code as well as write their own code to build upon existing programs, offering a novel approach to facilitating computational thinking skills through creative coding processes.

\subsection{LLMs for Creative Iteration}
GenAI has transformed the landscape of creativity, as tools like ChatGPT\footnote{https://chatgpt.com/}, Midjourney\footnote{https://www.midjourney.com/}, and RunwayML\footnote{https://runwayml.com} have demonstrated co-creative capabilities in writing, and image and video generation, prompting the invention of AI-powered creative platforms like FLORA\footnote{https://flora.ai} and ComfyUI\footnote{https://www.comfy.org}. These technologies have catalyzed significant interest in human-AI co-creativity, a collaborative process where both humans and AI work together as active partners to drive the creation of ideas and artifacts \cite{davis2013human}.

The emergence of human-AI co-creativity raises key questions about how such collaborations might be cultivated in educational settings. %, with recent work investigating how AI might support learners' ideation, experimentation, and reflection. 
\citet{omran2024redefining} introduced GenAI tools to courses in a creative technology program, showing that while AI may accelerate students' creative ideation, students reported experiencing choice paralysis from the abundance of potential ideas and expressed concerns that AI might diminish their creative agency. Educators and students in the arts valued GenAI as a means to overcome creative blocks but advise against overreliance that may hinder development of important creative skills \cite{saez2024analysing}. \citet{sandhaus2025co} further highlighted the importance for future GenAI to foster skills such as critical evaluation, ethical judgment, and human-centered design reasoning. \citet{situatedimaginariestas} studied computer science teaching assistants (TAs) and their perspectives on AI use, finding that while TAs use AI tools for grading and feedback, they also worry that AI may diminish connections between TAs and their students, highlighting the tension between ``[preserving] and [eroding] the human dimensions of their work.'' Together, these studies suggest that GenAI has significant potential to support creative ideation and exploratory workflows, while underscoring the need for human mentorship, carefully designed tools, and thoughtful pedagogical scaffolding. %to drive the development of foundational skills in creative thinking.

A growing number of LLM-based tools has been proposed to support creative exploration using creative code \cite{angert2023spellburst, wang2025pinning, keyframer, tseng2024keyframer, liu2025logomotion}, but have not centered on audiences \textit{learning} to write creative code themselves. Keyframer \cite{keyframer, tseng2024keyframer} supports design exploration for CSS-based animations by allowing its users to refine animations via prompting, code editing, and property controls. Spellburst \cite{angert2023spellburst} introduces a node-based environment supporting both natural language prompts and direct code edits, studying how expert creative coders can rapidly generate and iterate on creative code. Wang et al. \cite{wang2025pinning} demonstrate that decomposed prompting encourages reflection, goal revision, and strategic iteration among creative coders, allowing them to evaluate LLM-generated suggestions and make informed creative decisions. % Scratch Copilot \cite{druga2025scratch} provides youth learners with AI-assisted scaffolding through the creative coding process by assisting with ideation, debugging, and asset creation.
Building on this prior work, Flowcode explores how AI-assisted scaffolding can support creative ideation and iteration for \textit{new creative coders}. %Designed around example-based remixing, the Flowcode interface encourages students to focus on remix existing code. 
With Flowcode, users rely on their own creative agency to drive the direction of their projects, iterating upon the source code using  AI features designed to aid users in learning to actively write code themselves. % while FlowCode helps them realize their goals through a visual preview of what their idea might look like and scaffolded steps of how to implement it through code.
\section{Research Team Background}
\label{sec:positionality}

% more detail about type of experience teaching
% K-12, undergrad, grad
% in what capacity - TAing courses, Instructor 

Our research team draws from extensive computer science teaching and mentorship experience in both higher education and K-12 contexts. Six authors having taught undergraduate courses and workshops in User Interface Design, Data Science, and Introductory Computer Science; five authors have experience teaching K-12 programs such as game development and front-end development through organizations including Girls Who Code, Society of Women Engineers, and FIRST Robotics; and three authors served as peer tutors or advisors, mentoring students in introductory computer science. % , and facilitated workshops on data analysis, visualization, game design principles, and coding for beginners. 
% Our experiences have generally targeted beginners in various subjects, including efforts to demystify computer science as well as publishing research exploring how learners navigate information while working. 
This breadth of experience grounds our motivation for making computer science more accessible and informs our approach to how AI can support learners in creative coding contexts.

\section{Design Goals}
\label{sec:design-goals}

We designed Flowcode to help beginners iterate on creative coding projects, incorporating AI to encourage users to actively write code rather than solely use AI-generated code solutions. With this framing in mind, we established two central goals: \textbf{DG1}: Help learners understand existing code. \textbf{DG2}: Enable learners to productively iterate on existing code.

While there is a rich body of creative coding research in HCI centered on Processing and its JavaScript (JS) variant p5.js \cite{subbaraman2023forking, mcnutt2023study}, especially in connection with broadening participation \cite{wood2016computational, xu2018updating, xu2016creative, lovell2021craft}, we chose to examine a relatively under-examined but popular practice of creating animations using HTML, CSS, and JS. We chose to do this because (1) understanding how different web languages interact is a transferrable skill that is not normally learned when using a single language like Processing or p5.js; (2) different than Processing art, where designs are abstract and typically take on a mathematical and algorithmic approach, popular JS animation libraries like GSAP\footnote{https://gsap.com/} and Anime.js\footnote{https://animejs.com/} utilize imported assets like SVGs, lending itself to illustrated output instead; (3) there are established online communities for this type of creative coding such as CodePen\footnote{\url{https://codepen.io/}}, where over 10 million creative code projects have been shared ranging from front-end interfaces, CSS art, and animations. 

In this work, we focus on web-based animations created with the popular animation library Anime.js, used by both hobbyist and professional front-end developers. Part of our goal in selecting this approach is to shed light on the unique challenges that come with creative coding when working across HTML, CSS, and JS rather than a single language, and we hope to encourage others in the HCI community to consider this rich community of practice.
\section{Iteration 1 \& Pilot Workshop}

% We designed the Flowcode user experience drawing from our own backgrounds as computing educators. %along with incorporating feedback from several professors who teach creative computing at the undergraduate and graduate level (a detailed analysis of educator feedback is out of scope for this paper and in preparation for a separate publication). 
In this section, we describe the initial design of Flowcode along with a pilot study where we observed how students new to creative coding utilized the editor. We end with a summary of study results and how they informed our refinement of Flowcode's AI features.

\subsection{Initial System Design}

\begin{figure*}[htb]
    \centering
    \includegraphics[width=\textwidth]{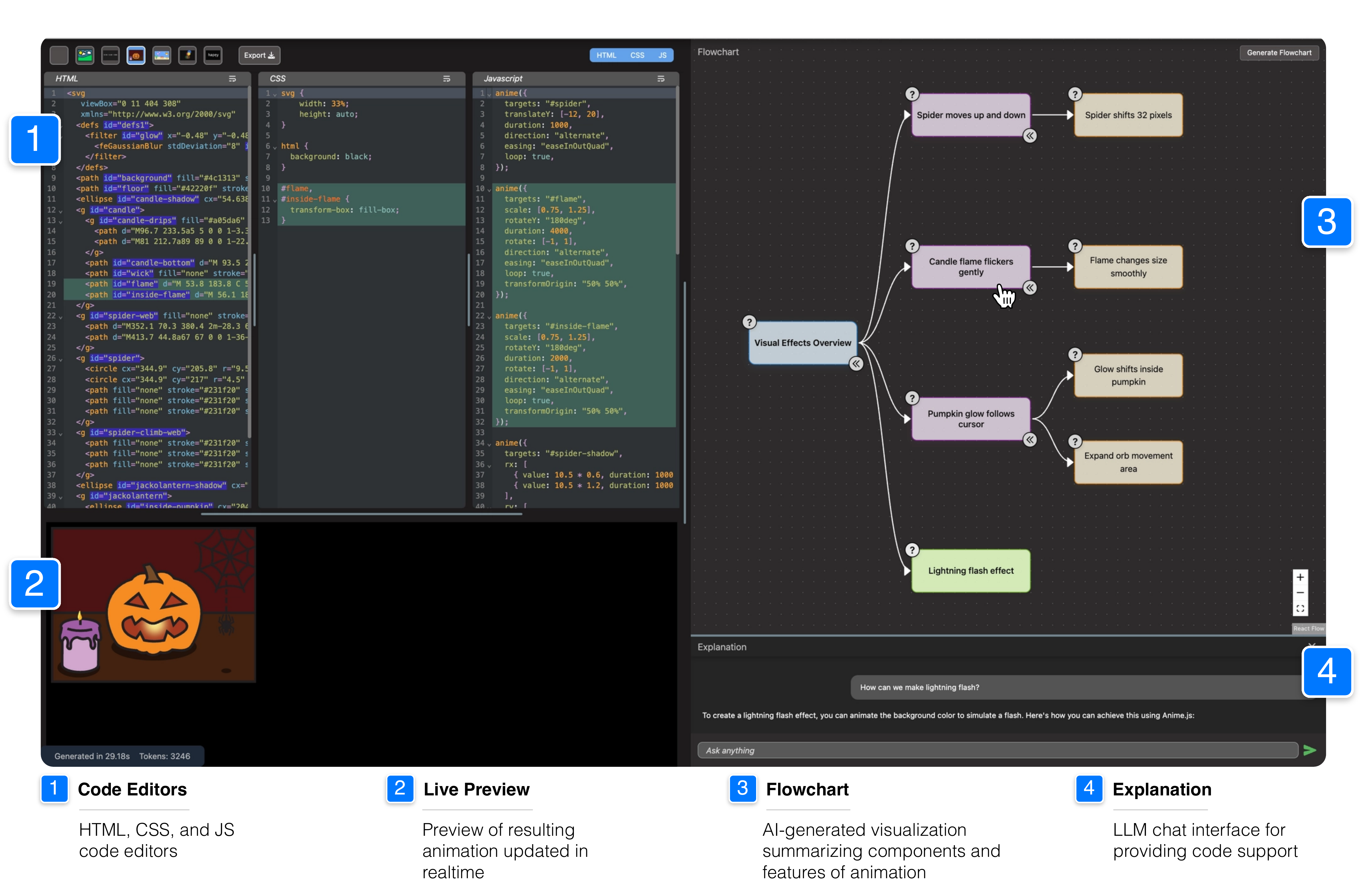}
    \caption{The Flowcode programming environment}
    \Description{} %this is where the alt caption for accessibility goes
    \label{fig:flowchart-system-diagram}
\end{figure*}

Flowcode’s user interface is partitioned into 4 major components: (1) the code editors, (2) the live preview, (3) the flowchart, and the (4) explanation panel as displayed in Figure \ref{fig:flowchart-system-diagram}.

\subsubsection{Code Editors \& Preview} In the (1) code editors, users write HTML, CSS, and/or JS to edit and extend a selected project\footnote{Users can select from existing projects, or starters, we have brought into the app or paste their own code to edit any project}. They can interact with the animation and view its design using the (2) live preview, which is automatically updated in real time whenever the code is edited (DG1).  

\subsubsection{Flowchart} The (3) flowchart provides a visual summary of the project's elements and links these summaries to lines in the code. The flowchart's content is generated by an LLM given the project’s code as context. Our prompt requests a beginner-friendly summary of the project's main features across two levels of abstraction: the primary level (purple nodes), which describes distinct features in the design; and the secondary level (yellow nodes), which provide high-level technical descriptions of how each feature is implemented using keywords and values in the animation code. 

For example, Figure  \ref{fig:flowchart-system-diagram} displays an example of a Halloween animation with three main components: a spider that moves up and down, a flickering candle flame, and a pumpkin whose inner glow moves with the mouse cursor. The flowchart distills the animation into these primary components (e.g., \textit{Spider moves up and down} and \textit{Pumpkin glow follows cursor}), with the secondary nodes reflecting the technical implementation of each component (e.g., \textit{Spider shifts 32 pixels} and \textit{Glow shifts inside pumpkin} along with \textit{Expand orb movement area}). Our prompt guides the LLM towards generating beginner-oriented descriptions as node labels and distinguish between primary (tier 2) and secondary nodes (tier 3), with an excerpt shared below (the full prompt is in Appendix \ref{app:flowchart-prompt}):

\begin{shaded}
\begin{lstlisting}[frame=none, basicstyle=\small] 
Given the following code for a creative coding project, generate a JSON flowchart that represents its structure and logic. This flowchart should be geared toward helping beginner coders understand the VISUAL and INTERACTIVE effects of the code.
...
TIER 2 GUIDELINES:
- START WITH ACTION VERB: "Drift", "Spin", "Fade", "Bounce", "Scale", "Rotate"
- SPECIFY THE TARGET: "the clouds", "the petals", "the background", "the dots"
- ADD DIRECTION/MANNER: "leftward", "gently", "to dark", "upward", "larger"
- KEEP IT VISUAL: Focus on what the viewer sees happening, not code mechanics
...
TIER 3 NODE GUIDELINES (expand on parent with concrete, code-informed detail -- NO code syntax):
- PURPOSE: Each tier-3 node answers "HOW does the parent's visual effect actually work?" by translating specific code parameters into plain, beginner-friendly language. It must teach the reader something NEW that the parent node did not already say.
- REFERENCE CODE PARAMETERS WITHOUT SYNTAX: Describe what the code controls -- specific pixel distances, durations in seconds, color names, directions, repetition behavior -- but ALWAYS in everyday language. NEVER use code notation like property names (translateY, opacity, scale), selectors in quotes (`.dot`), colon-separated key-value pairs (duration: 2000), or function names (easeInOutSine).
- EXPAND, DON'T REPHRASE: The parent describes the overall visual action. The child must zoom into a SPECIFIC mechanism, value, or behavior that makes that action work. Test: "If I removed this child node, would the reader lose understanding of HOW the parent's effect happens?" If the answer is no, the node is redundant and should be rewritten.
\end{lstlisting}
\end{shaded}

Web-based creative coding projects are typically implemented across multiple languages and files (HTML, CSS, and JS). Traversing files to understand how and where features are implemented can be a challenging task for beginners. To overcome this hurdle, users can click on nodes in the flowchart to highlight relevant code across all three editors, which shows the relationship between files (e.g., where in the HTML an image is defined, how it is styled in CSS, and how it is animated in JS). In the example shown in Figure \ref{fig:flowchart-system-diagram}, clicking on the primary node \textit{Candle flame flickers gently} highlights the corresponding code for the candle flame across the HTML, CSS, and JS code. These highlights support users understanding the implementation of the animation by connecting semantic descriptions to individual code fragments (DG1). For full details about the prompts use to construct the structure of the flowchart, along with handling code highlighting from flowchart nodes, please refer to Appendix \ref{app:flowchart-prompt}.

\subsubsection{Explanation} 
\begin{figure}
    \centering
    \includegraphics[width=0.4\textwidth]{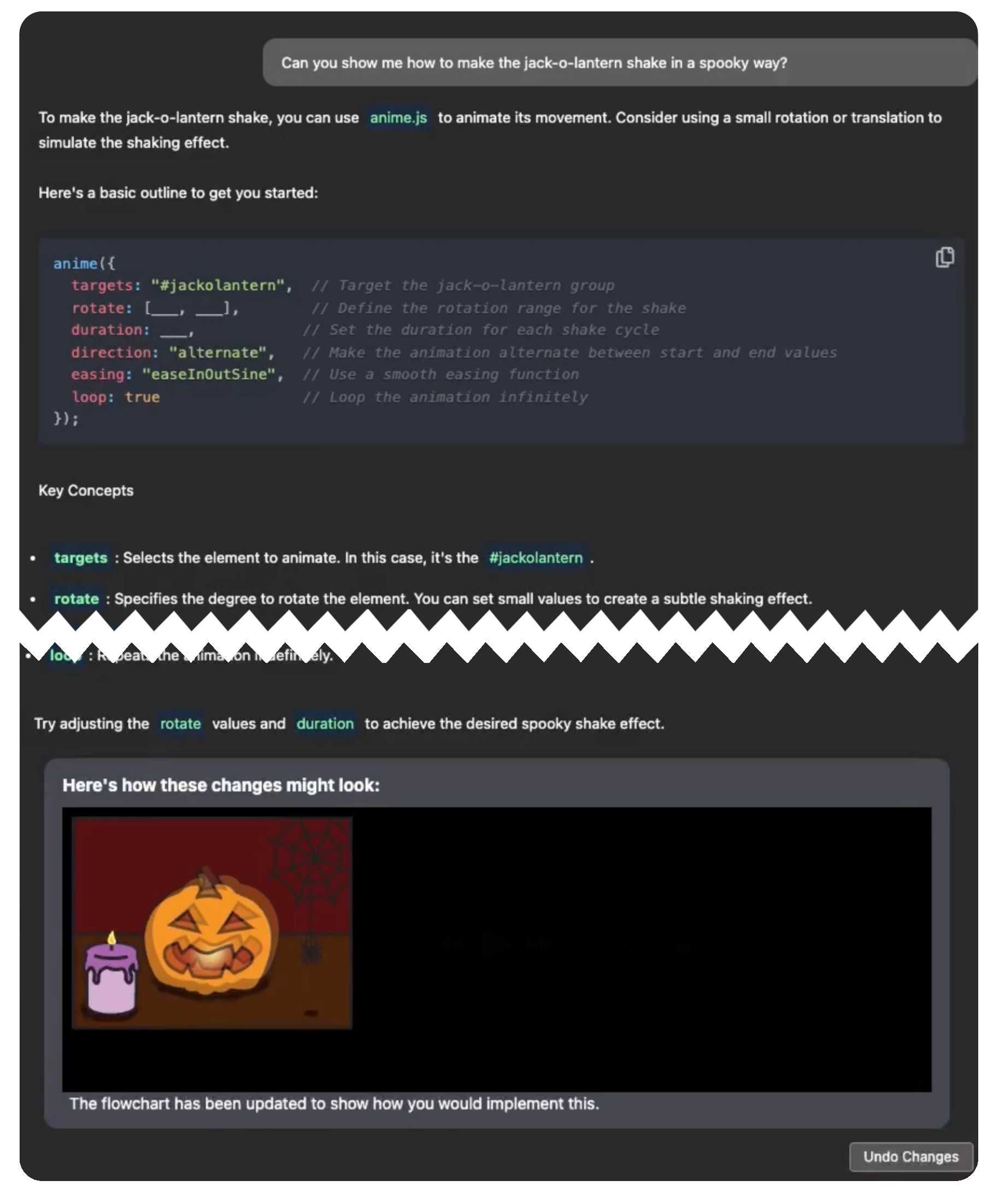}
    \caption{Generated explanations in Flowcode display fill-in-the-blank code snippets and previews of what the feature would look like when implemented.}
    \Description{} %this is where the alt caption for accessibility goes
    \label{fig:explanation}
\end{figure}

Alongside the flowchart, we implemented an (4) explanation panel where users can prompt an LLM for guidance. The LLM was tailored to provide beginner-oriented explanations and encourage learning and understanding while minimizing vibe coding. Our team recognized that when students use chat interfaces like ChatGPT to ask about code, LLMs typically return full code solutions alongside lengthy explanations. We combat this issue by prompting the LLM to provide shorter, beginner-oriented explanations and to produce \textit{fill-in-the-blank} code snippets (DG2) rather than providing a full solution as shown in Figure \ref{fig:explanation}. The blanks are designed as scaffolding for learners to write critical parts of the code themselves. Explanations with code are also incorporated into an interactive preview that presents the resulting animation if the user successfully integrates the code snippet, as shown in Figure \ref{fig:explanation}. This helps learners verify if the LLM’s explanation aligns with their design vision before implementing the feature. % In this prompt, the adjective `realistic' is abstracted from Anime.js properties one might use to achieve this effect, with the LLM suggested animating the fill and opacity properties of the flame and displaying the result in the explanation preview as shown in Figure \todo{#} \todo{FIGURE: show the fill in the blank code and preview}. 

When generated explanations contain code, the system automatically updates the flowchart and code to represent the new feature. This includes adding a new node describing the feature in the flowchart, inserting the code snippet with blanks into the corresponding code editor, and updating the line number references in the flowchart to maintain correct line references with existing nodes. Users can revert these changes with an undo button, and they can also regenerate the flowchart manually if needed.

Users can provide specific context alongside their prompt by clicking a question icon that appears when a user highlights a line of code or alongside each node in the flowchart, which passes the code selection or node title to the LLM alongside their query.

Putting these features together, the user is able to utilize both the flowchart and explanation panel for guidance as they iterate on an animation.

\subsubsection{Technical Implementation} Flowcode is a Node-based web application that utilizes GPT-4o as its underlying model. The flowchart is implemented with React Flow\footnote{https://reactflow.dev/} and we use the LLM to generate JSON representing the flowchart content. Please refer to Appendix \ref{app:prompts} for details about our prompts.

% % refactor below
% At the time our pilot workshop took place (November 2025), Flowcode differed from the design described in our Systems Description (Section \ref{sec:system}) in several key ways:
% \begin{enumerate}
%     \item \textbf{Consolidated Explanation} - While our explanations at this stage were designed to utilize fill-in-the-blank code snippets rather than complete code, and to produce concise beginner-oriented responses, the explanation was shown completely at once (not step-by-step), each starting with a textual explanation and ending with a fill-in-the-blank code snippet \todo{include screenshot}.
%     \item \textbf{Auto-Updating Flowchart and JS Code} - Each time a user prompted the LLM with a question about implementing a new feature, Flowcode would \textit{automatically} update their Flowchart with a new node describing the feature (with the option for the user to undo this change). In order for the newly added node to correctly highlight lines in the user's code editors when clicked, we also auto-injected the fill-in-the-blank code snippet and accompanying comments into their JS editor.
% \end{enumerate}
% Our results in this section will elaborate on the ways our decisions to revise these features were informed by how we observed Flowcode being used in this pilot.

\subsection{Study Methodology}
\label{sec:pilot-methods}

% In this section, we describe a workshop we designed for beginners to build animations in Flowcode, describing what we discovered about how beginners used and perceived Flowcode's AI features, and how these results informed our refinement of the editor, leading into a second iteration described in the Section \ref{sec:1-1-study}. The results from these two studies ultimately enable us to compare and reflect upon how our changes impacted the role AI played in our users' iterative design processes.

We designed a pilot workshop for beginners to build animations in Flowcode. Student participants for this workshop were recruited at a private research university in the Northeastern United States. Recruitment emails were sent to various mailing lists on campus geared toward design and computing interest groups. We selected 10 participants from the 31 students who filled out our screener survey, selecting for diversity among years and majors and choosing participants who self-reported having beginner or moderate experience with HTML and CSS and beginner-level experience with JS. We chose this workshop size because we used a computer lab on campus that contained 10 desktop computers and monitors, and we wanted to keep the workshop small enough that our research team could observe all participants easily. Ultimately, seven students attended and completed the workshop. The seven students consisted of 4 undergraduate and 3 graduate students, representing a variety of majors including computer science, economics, astrophysics, data science, and urban planning. % Five out of seven responded to our demographics survey, all identified as female. % computer science (3/7), economics (2/7), astrophysics (1/7), data science (1/7), and urban planning (1/7)}. 

The workshop began with a 20-minute facilitated introduction to Anime.js and the Flowcode editor. %, walking through building a simple SVG-based animation from scratch to demonstrate the syntax of Anime.js and how it can be used to target HTML and CSS elements to animate individual properties. 
Next, each participant used Flowcode in three design activities lasting 15, 30, and 45 minutes respectively. In each activity, participants built on top of starter projects our research team created that utilize SVG images and text to animate. % All participants animated the same two projects for the first two activities, and were then free to pick from a set of four starter animations. 
For all activities, participants were free to choose what design features they wanted to add and to utilize Flowcode features in whatever way supported their process. Four researchers observed 1-2 participants each throughout the workshop and took observation notes. At the end of the workshop, participants provided feedback on their experience using a collaborative Miro board and group discussion. Each participant received a \$50 gift certificate and free lunch. This study was approved by our institution's IRB.

Data collected during the workshop included screen recordings of participants' interactions with Flowcode, researcher observation notes, Miro board feedback, and interaction logs collected in the Flowcode app, which includes logged events such as user prompts and code edits. We analyzed the data qualitatively using thematic analysis \cite{braunUsing2006} to identify themes in participants’ feedback, and we examined participants' flowchart usage and prompting approaches using the interaction logs and screen recordings. A group of five researchers reviewed and analyzed the data inductively over the course of two months.

\subsection{Results}
\label{sec:pilot-results}
Participants in the workshop were all introduced to creating animations with Anime.js, with three choosing to remix an animated Halloween scene and four remixing a particle system animation in the third activity. On average, participants prompted the LLM 32 times during the workshop (\textit{SD}=8.2), using the it frequently to ask for assistance implementing new features, answer general programming questions (like color HEX codes), and provide help debugging their code. While participants' shared positive feedback about the flowchart, the workshop also surfaced technical and usability issues that ran counter to our design intentions, which we describe in the following section. These issues led us to refine the Flowcode prototype and several elements of our study design for our second iteration.

\subsubsection{Fill-in-the-Blank Code Generation Issue} We unexpectedly encountered a significant issue with the quality of the LLM's responses, where code snippets generated by the model did not reliably utilize the fill-in-the-blank approach and instead produced complete code suggestions. This behavior varied among participants, with two of the seven participants never encountering the problem, whereas one participant saw 24 of their prompts incorrectly return full code snippets. Notably, once the LLM incorrectly generated complete code for a specific user, it continued to do so in most subsequent responses throughout the workshop. As we are missing part of a screen recording for one participant, this precludes us from quantifying the frequency of this issue for all users, but the other 4 users saw between 4-6 complete code snippets. Our team significantly reworked our prompting strategy for code explanations in response to this issue. Another problem that arose was that the LLM did not always strategically place blanks in code snippets; for example, it sometimes created blanks for trivial parameters like duration (the length of time an animation should take place), while other more critical parameters for the animation would have generated values. (In Section \ref{sec:explanation-prompt-refinement} we describe refinements to our prompt). Despite this core issue with generated code snippets, we were still able to observe the ways participants interacted with the flowchart, which we highlight next.
% On average, the number of full code snippets generated in our Explanation interface was 24\% (\textit{SD}= 23\%). 

\subsubsection{Flowchart Use}
Participants utilized the flowchart in several distinct ways. First, the flowchart assisted in \textit{orientation}, where learners explored the structure of the code and where key features were implemented before editing themselves. Three of the seven participants began by clicking on nodes in the flowchart before prompting the LLM or editing the code. Participant WP7 described this process by stating, ``It helped me think through the order of steps and which code blocks are nested within what step. It was easy to navigate such that I can see the relevant block [of code] highlighted.'' WP3 shared how this made the project she remixed easier to follow: ``Flowcode helps visually segregate the different moving parts, which may help breakdown a large project into digestible parts.'' Moreover, WP2 and WP6 expressed that the flowchart helped them understand the \textit{relationship} between different animation effects due to the organization of the flowchart and which features were nested under which primary node. WP4 discussed how it supported her understanding of ``what processes were running parallel and what functions had other dependencies,'' while WP2 shared, ``The flowchart was helpful in summarizing what the code was doing and the order it was being done,'' indicating that the flowchart assists with understanding sequencing.

Further, the flowchart was used for \textit{navigation}, where participants interacted with the flowchart by clicking on nodes (4/7) or zooming into or panning on the flowchart (7/7) \textit{while} working, suggesting continual monitoring of their code. %WP5 shared, ``"The flowchart was helpful in finding the corresponding lines of code to a certain task/step." 
For two participants, navigation actions were directly connected to \textit{verifying} their code, where they utilized manual regeneration of the flowchart to verify changes. While working, WP4 clicked on a node related to a feature she no longer wanted to be part of her animation, then subsequently deleted the highlighted code from her editor. She then regenerated the flowchart to verify that the functionality had been successfully removed. Similarly, WP2 regenerated the flowchart several times after making large manual changes to her code; the regenerated flowchart was much smaller, which accurately represented the changes she made.

\subsubsection{Drawbacks to Automating Updates} At this point in time, the flowchart was designed to automatically update each time a user's prompt to the LLM produced code suggestions (for example, when a user asked for help implementing a new feature in their animation). In effect, new nodes were added to the flowchart in response to user prompting. To ensure line highlighting was correct when users clicked on existing nodes, we automatically added the generated code snippet (often with fill-in-the-blanks) into the code editors and regenerated the flowchart to ensure consistent line references. 

However, this proactive approach led to several unintended consequences that we felt negatively impacted the learning experience. First, the flowcharts became quite large as users worked on their project, with new nodes frequently being appended. This made the flowcharts more difficult to follow the longer the project was worked on, a usability issue that seemed aligned with participants seeming to passively monitor changes to the flowchart rather than actively interact with it after these automatic updates. From a practical standpoint, this also caused the creation of many LLM requests for details users didn't always require, which had cost implications.

Additionally, alongside nodes being added to the flowchart, the code editors would be updated with the newly generated code by scrolling to and highlighting the added code to help users identify where additions were made. However, this visual change drew participants to the code and consequently away from the generated explanations; they went straight to editing rather than r®eviewing accompanying explanations. Similarly, we saw that when interacting with generated explanations, participants would often scroll to the code snippet at the bottom of the explanation, suggesting there was limited review of the explanations when code was presented.

\subsubsection{Takeaways} Overall, we found participants' feedback about the flowchart promising, while we also identified technical issues with generated LLM code suggestions and limited review of explanations resulting from our automatic updates to the flowchart and the code. This made our team reconsider how we might better structure explanations to deepen learner engagement, as well as enable more intentional flowchart updates that keep the visualization readable as users iterate on their designs. 

\section{Iteration 2 \& User Study}
\label{sec:1-1-study}

Informed by our pilot, we made significant updates to our explanation interface and flowchart updates, with the goal of supporting learner engagement with generated explanation and code. In this section, we describe our updated interface along with a follow-up user study we conducted to test the effect of these changes.

\subsection{System and Study Updates}
% We made several refinements to Flowcode in an effort to improve engagement with generated explanations. To evaluate these changes, we ran a lab study in which participants could share their process they created their animations. 

\subsubsection{Step-by-Step Explanations}
\begin{figure*}[htb]
    \centering
    \includegraphics[width=0.9\textwidth]{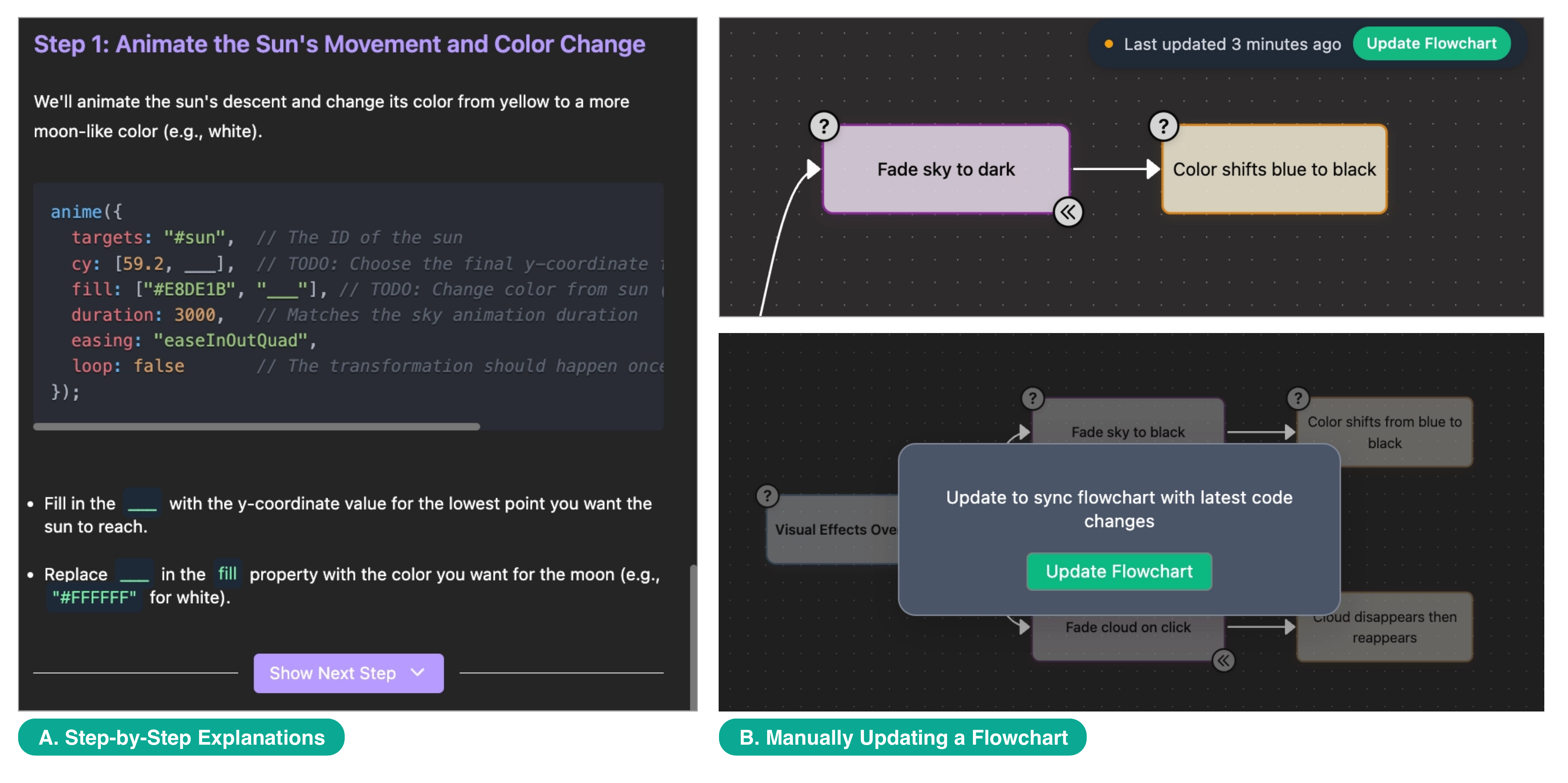}
    \caption{(A) Explanations broken into steps. Users must click the `Show Next Step' button to progress to the next step, rather than viewing the entire explanation at once. (B) Users see a notice when the flowchart can be updated. If they attempt to click on a node when their flowchart is out of date, we ask them to update their flowchart, which ensures line references are correct.}
    \Description{} %this is where the alt caption for accessibility goes
    \label{fig:step-by-step-update}
\end{figure*}

In our pilot, we observed that when LLM responses consist of both natural language explanations and code snippets, users were likely to skip over the explanation to go directly to the code, leading to minimal review of the explanation. To combat this, we updated the structure of the explanation so that the LLM breaks down explanations into individual steps that are not all exposed at once. Instead, the user must explicitly click a button to reveal the next step, adding a layer of friction to encourage users to read the explanation before progressing as shown in Figure \ref{fig:step-by-step-update}A. Suggested code changes may also be spread across multiple steps rather than shown in full at the end of an explanation, reducing the possibility of scrolling to a solution. 

\subsubsection{Removed Auto-Updating of Flowchart and Code} 
Previously, when the LLM returned code suggestions as part of its response, we automatically added new node(s) to the flowchart and injected the generated code snippet into the code editor. This had several unintended side effects: (1) users would gravitate to the added code snippet in the editor and start editing with minimal interaction with the generated explanations, %(2) the flowchart updates were seen as ambient, which we thought may reduce the likelihood users would engage with it, 
and (2) the flowchart would grow in size after each update, which reduced readability. 

In response, we removed auto-updating of the flowchart and instead have messaging to let users know when it is out of date, with an option to manually regenerate it as shown in Figure \ref{fig:step-by-step-update}B. % If a user attempts to click on a node in an out-of-sync flowchart, the user must update the flowchart before continuing so we can ensure line highlighting works correctly with a user's most recent code edits. 
This change also affects how users interact with code from explanations -- rather than automatically adding the code snippets to the editors, users need to manually copy and paste select lines over as they review the explanation. We removed a copy-to-clipboard button previously available next to code snippets to try to minimize copying without reviewing. Users thus need to carefully consider where the generated sample code should be placed into their design, along with what values need to be added for any blanks.

\subsubsection{Explanation Prompt Refinement}
\label{sec:explanation-prompt-refinement}
Our prompt for generating explanations was greatly refined from our pilot workshop, changing from a 314-word prompt to a 2724-word prompt. First, we provided guidelines for how the LLM should consider placing blanks in code snippets to be more pedagogically useful while asking the LLM to provide default values for particular properties like duration and easing, which were not substantively linked to the types of questions we saw learners ask about implementing new animation features. We also shared both correct and bad examples of code snippets for all languages (HTML, CSS, and JS) to provide concrete examples of poor blank placement as well as incorrect fully complete code snippets. Below, we provide an excerpt for our prompt to share some of the pedagogical guidelines we provide for the LLM to consider the placement of blanks.

\begin{shaded}
\begin{lstlisting}[frame=none, basicstyle=\small] 
You are a virtual TA within a beginner-friendly creative coding environment using HTML, CSS, and JavaScript with Anime.js, similar to CodePen. Your goal is to foster critical thinking and problem-solving, **not** provide direct answers or complete code or encourage vibe-coding. 
...
In response to questions about code, you will create an explanation that guides a beginner towards answering their question.  Your explanation MUST incorporate code that is INCOMPLETE and requires users to fill in blanks that strengthen their understanding. You are NOT just formatting code with blanks - you are designing a learning experience that supports how people learn to code.
...
Step by step explanations should NEVER contain full code snippets and should instead have pedagogically placed blanks so users have to write code to complete the feature. When deciding where to place blanks in your code explanations, FIRST ask "What is the hardest concept here for a beginner," then "What makes this hard for beginners"? Design blanks around this specific concept. Each blank should:
- Force the user to apply a concept, not just fill in a trivial value
- Build toward mastery of the underlying programming pattern 
...
PRIORITIZE blanks that involve these important programming concepts, in this order (most important first):
1. **Core logic/algorithm** - loops, conditionals, function calls with parameters
2. **Key property values that change behavior** - selectors, transform values, array/object structures
3. **Conceptual connections** - linking variables, understanding scope, using return values
4. **Configuration details** - duration, easing, timing values (these can have defaults)

If code has:
  - Array/object manipulation - blank the structure
  - Function calls - blank the parameters
  - Logic patterns - blank the conditions/expressions
  - DO NOT just blank the configuration values, which makes the programming exercise trivial

REQUIREMENTS for fill-in-the-blank scaffolding:
- **DO NOT provide complete code blocks a user can copy and paste**
- Use either ___ for missing values or // TODO comments for missing logic.
- **NEVER provide the values for blanks** in comments or in explanations. Comments should help guide the user and not provide the solution.
- **NEVER use blanks for trivial values** - If the hardest blank is a single number or string with no conceptual weight, you've failed
- PICK LOGICAL DEFAULTS for 'duration' (1000) and 'easing' properties - DO NOT use blanks here.
\end{lstlisting}
\end{shaded}

Full details on our prompt can be found in Appendix \ref{app:iteration-2-prompts}.

\subsubsection{Updates to Study}
Reflecting on our study format from our pilot workshop, we felt that while we saw how users interacted with the editor, it was difficult to elicit \textit{why} they chose to use the editor in particular ways in a group setting. As a result, we decided a lab study in which we can ask individuals to think aloud while they work would help us better link their motivations and goals to their actions. %Finally, we implemented more detailed logging , such as when users pan in the flowchart interface and the full generated responses from the LLM so we could more easily evaluate the quality of the LLM's responses.
Additionally, for the pilot workshop, our team created the set of animations participants built on top of, which were purposely well formatted and organized while utilizing features of Anime.js that were more beginner-oriented. While this made sense in the context of an educational workshop, we thought it may be less representative of examples users might find online. Thus, we decided to instead present users with actual animation code from publicly shared projects on CodePen, aligning with how one might learn from examples on the web. Other than light resizing of animations (e.g., editing width and height parameters in CSS) to ensure the animation could be fully visible in Flowcode's live preview, the code was left untouched. Full details about the selection of these Anime.js CodePen examples can be found in Appendix \ref{app:codepen}.

\subsection{Lab Study Recruitment Details}
\label{app:lab-recruitment}
For our follow-up lab study, we recruited participants by emailing students who expressed interest in our pilot workshop but were not selected, and through snowball sampling by asking these students to forward the opportunity with others who were eligible. We considered participants eligible if they had prior experience with HTML and CSS but were beginners to JS or new to creative coding.

\subsection{Study Methodology}

We designed a lab study to understand how novice creative coders used this updated version of Flowcode. We had 9 participants who were all female, predominantly undergraduates (8/9) with one graduate student, studying majors ranging from computer science, cognitive science, education, and mechanical engineering. For full details about our recruitment, please refer to Appendix \ref{app:lab-recruitment}.% 

Participants used Flowcode in 90-minute sessions over Zoom, where they iterated on JS-based animation brought into Flowcode from CodePen. Sessions began with a five-minute background interview asking about the participant's programming experience and experience using LLMs for coding tasks. Then the facilitator gave a 10-minute introduction to Anime.js and the features of the Flowcode editor. The participant remixed a simple starter in a 15-minute activity to familiarize themselves with the editor, followed by the main task of picking from four CodePen animation projects and iterating on them in whatever way they wanted over 40 minutes. We asked them to target adding 2-3 new features to the design and think-aloud while working. We concluded with a 10-minute semi-structured interview asking participants to reflect on their experience. Participants received a \$45 gift card for completing the study. This study was approved by our Institution's IRB. 

We collected the following data: (1) recordings of each Zoom session (13.5 hours of video), (2) video recording transcripts, (3) instrumented logs of user interactions such as when users regenerated the flowchart, made code edits, and prompted the LLM (12k logs), (4) researcher observation notes from the session, and (5) researcher memos written immediately after the study sessions that captured researcher reflections. We analyzed the transcripts alongside screen recordings using a thematic analysis approach to identify common use patterns and themes \cite{braunUsing2006}. We analyzed log data to examine the quality of the model's generated responses and create visualizations showing how users worked across multiple features of Flowcode as they iterated on projects.

\subsection{Results}
In the following sections, we focus on participants' interactions with the editor during the 40-minute second activity. %, since the first activity's primary focus was familiarizing users with the editor. 
We begin with a vignette to illustrate how one participant iterated on their project utilizing all of Flowcode's features, followed by a summary of all participants' interactions, and concluding with two sections focusing on use of the flowchart and use of explanation features.

\subsubsection{User Vignette: Iterating with Flowcode}

\begin{figure*}[htb]
    \centering
    \includegraphics[width=\textwidth]{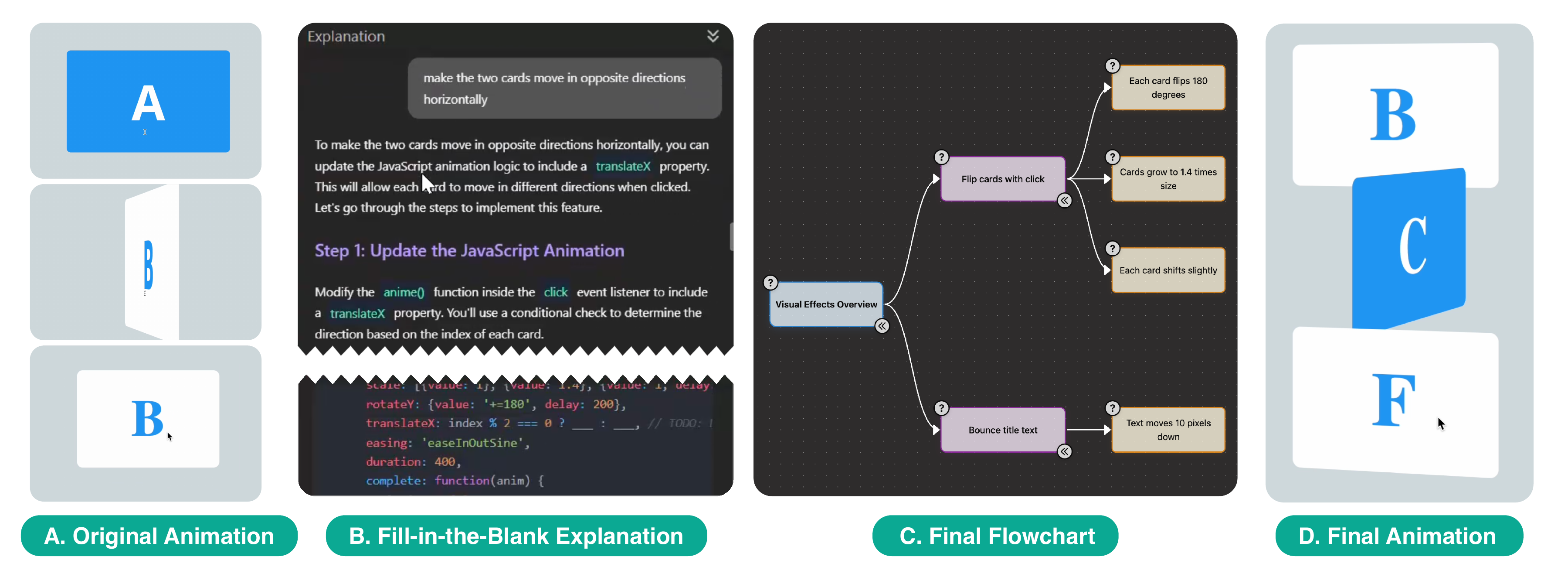}
    \caption{Stages of P1's process of remixing a card flip animation. }
    \Description{} %this is where the alt caption for accessibility goes
    \label{fig:vignette-flowchart-orientation}
\end{figure*}

Our vignette describes P1, a computer science graduate student new to creative coding. % who self-described as having beginner-level experience in HTML, CSS, and JS %drawn from working on front-end web design for academic projects, but with no prior creative coding experience. 
She chose to work on the Card Flip starter, which features a double-sided card that animates to reveal the opposite site when clicked on, as shown in Figure \ref{fig:vignette-flowchart-orientation}A.
% with the letter `A' written on the front and the letter `B' written on the back. Clicking on the card animates a card flip to reveal the opposite side, scaling the card and up and down during the animation as shown in Appendix \ref{app:codepen}. % We selected P1 for our vignette because she interacted with many of Flowcode's features and created an animation that was fairly distinct from the original design. 
% To begin, P1 focused on reviewing the existing HTML, CSS, and JS code by reading and clicking on nodes in the Flowchart. %In Figure 1, we present one such interaction. 
To start, P1 was curious about what parts of the code caused the card to scale up and then back down when flipping, so she explored the flowchart and clicked the node `Card scales up then back' and reviewed the HTML/CSS/JS code it highlighted. She also edited the card rotation angle from 180 degrees to 360 degrees in JS to see how this would affect the animation. 

After her initial orientation, she started using the Explanation feature to implement her design ideas. She entered the prompt \textit{Add another card with the same flipping features for the letters C and D.} In response, the LLM generated a multi-step explanation with fill-in-the-blank code and an interactive preview. %The first step described the HTML structure for adding a second card, and the second described the adjustments to the JS that would allow interacting with the new card. 
Before editing the code, she read through the first step, expanded and read the second step, and clicked on the explanation preview to test how these changes would look. The explanation included a blank and comment about editing an Anime \textit{targets} property, asking her to consider which HTML element should be targeted. After her first edit, she tested using the live preview and found that only the first card would flip. The live preview provided real-time feedback that allowed P1 to quickly identify the bug, re-read the explanation, and update the variable to select all cards.

Next, she entered the prompt \textit{make the two cards move in opposite directions horizontally}. As shown in Figure \ref{fig:vignette-flowchart-orientation}B, the LLM suggested adding the \textit{translateX} propery, allowing P1 to experiment with the movement of each card. Doing so created a different animation from what P1 had envisioned; while her idea was to have the cards moving back-and-forth horizontally across the screen, the LLM's version translated the horizontal position of each card while also being flipped. %Because it was only necessary to change one line to achieve this, the LLM returned a single-step explanation with fill-in-the-blank code, allowing P1 to experiment with the movement of each card by editing the \textit{translateX} property. The LLM's suggestions did not align with P1's vision. Her idea was to have the cards move horizontally back and forth across the screen. \todo{This is my best guess as to what she intended, judging from the prompt and the video. I'm not sure, please feel free to verify: 51 minutes + 81 minutes into the video}. However, the LLM's version translated each card in the horizontal direction upon each flip. 
% During the study, she suggested that perhaps her prompt was too vague yet also noted that she actually preferred the version the LLM provided.
P1 shared that she actually preferred this design; in this way, the LLM's explanation helped her realize an idea she had not considered.
%``In my mind, I was like, I want to move this across the screen \textit{[Moves cursor across screen to illustrate her original idea]}.  But then [the final result from the LLM] [...] was different, and I ended up liking that more." 
%In this way, ambiguity left her request up for interpretation, and the LLM's explanation helped her realize an idea she had not considered. 
Before finishing the project, she regenerated the flowchart as shown in Figure \ref{fig:vignette-flowchart-orientation}C which showed her addition of making `each card shift slightly.'

P1 tended to prompt the Flowcode LLM in an imperative way (e.g., asking the LLM to do things for her by starting her prompts with verbs like `add,' `give,' and `make'); regardless, Flowcode avoided giving away the direct solution and instead gave her code snippets she had to complete herself. % Explanations were typically multiple steps, %(unless the change was minimal), encouraging P1 to work on one aspect of the implementation at a time. The fill-in-the-blank structure allowed her to experiment with the values, and the lack of a copy-paste button prompted her to read through the code to identify which lines were relevant. % She verified the expected behavior of the code using the explanation preview, and she identified bugs in her code using the live preview.  
Notably, towards the end, we observed a shift in her prompting approach as she asked a general programming question (\textit{What other parameters can be used for `direction'?}). This change suggested that she was progressing to a more learner-oriented approach to prompting as she considered other values she can use to tweak the code herself. Additionally, she added a third card and tweaked the flipping animation without the assistance of the LLM. By the end of the activity, she made changes independently, relying on her previous experiences and understanding rather than the LLM.

\begin{figure*}[htb]
    \centering
    \includegraphics[width=0.75\textwidth]{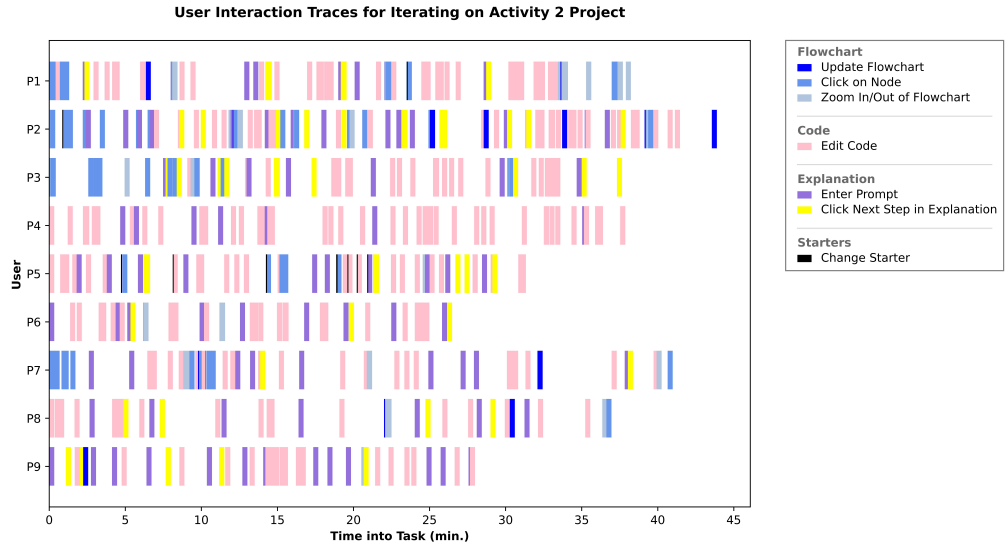}
    \caption{Interaction traces of how users navigated across features of the Flowchart editor.}
    \Description{} %this is where the alt caption for accessibility goes
    \label{fig:interaction-traces}
\end{figure*}

\subsubsection{Summary of Interactions}

Overall, we found that participants iterated on features using a combination of direct edits to code and support from AI features. When asked for their general feedback about Flowcode in our post-activity interview, 8/9 participants brought up on their own that they especially appreciate the integration of the Flowcode UI components: that the HTML, CSS, and JS code, along with the live preview and AI features, were all shown together in a single window, rather than having to go back and forth between windows (e.g., an IDE, a chat interface, and a tab for previewing the output of the code). The LLM's ability to generate fill-in-the-blank code snippets was more consistent compared to the pilot, while not being fully eliminated -- for more details about the quality of LLM generated fill-in-the-blank code snippets from this study, please refer to Appendix \ref{app:quality}.

Figure \ref{fig:interaction-traces} presents interaction traces of how individuals navigated across the flowchart, explanation interface, and code editors during their process. P7, P8, and P9 at the bottom of the figure represent three distinct interaction patterns for initially engaging with Flowcode: \textit{flowchart-first}, \textit{code-first}, and \textit{prompt-first} respectively.

P7's \textit{flowchart-first} approach is evidenced by the light blue `click on node` logs at the beginning of her trace. She began by inspecting the code with the help of the flowchart to understand the animation's SVG elements before attempting to add a unique SVG path on her own. Immediately after interacting with the flowchart, she entered two prompts (shown in purple) asking questions \textit{about} the code (e.g., \textit{what does d do in this code}). This flowchart-first approach symbolizes a user focused on iteration backed by understanding. As seen later in her trace, she continues to use the flowchart, regenerating it manually later (indicated by a primary blue bar) to ensure she can find segments of the code she wants to edit.

P8's \textit{code-first} approach demonstrates a user that began instead with attempting to implement a feature by editing the code directly. Her trace shows a few code edits (in pink) at the beginning, where she tried to rotate an SVG element. When she was unable to successfully add this change on her own, she then prompted the LLM for support with debugging. We liken this approach to interacting with the LLM like a TA after making their own attempt, using the assistant to help fix bugs. Aligned with this idea, her continued use of the flowchart later on was motivated by fixing issues in her code.

P9's \textit{prompt-first} approach highlights another type of user who begins by prompting the LLM to implement an idea (purple). She continually alternated between prompting the LLM, clicking through steps in the explanation, and editing code. This approach represents a user who relies on AI to accelerate toward their design vision, particularly since all new features she attempted started with a prompt, rather than trying to implement herself first. A user who takes a prompt-first approach may place higher emphasis on the ultimate design outcome than fully understanding the code.

Broadly, these three interaction patterns illustrate ways learners approached iterating on their code depending on their goals. In our study, 4/9 participants took a flowchart-first approach, 3 took a code-first approach, and 2 took an LLM-first approach. While these categories reflect how one might initially engage with the editor, users worked across these approaches as they iterated, largely using the explanation feature and flowchart in conjunction to aid with implement new ideas and debugging their code. 

Participants prompted the LLM between 6-15 times while iterating on their project (\textit{M}=9, \textit{SD}=3), with the most frequent prompts asking for code explanations (\textit{M}=3), followed by asking for support implementing a new idea (\textit{M}=2) and refining their design (\textit{M}=1.9). Other prompt types included asking general programming questions not specific to the user's project (e.g., \textit{What are types of easing values I can use?}) and help debugging their project. In other words, participants used the LLM to support all stages of their design process, from explaining existing code, debugging, and implementing new ideas. Participants primarily made JS-based edits to their animations (\textit{M}=16.9) along with a similar number of HTML (\textit{M}=7.1) and CSS (\textit{M}=6.1) edits, supporting the importance of being able to effectively navigate across different web programming languages when realizing a design, a feature we support through code highlights using the flowchart.

\subsubsection{Flowchart: Understanding Code Structure and Relationships} All but one user (8/9) utilized the flowchart during the study (by clicking on nodes or panning/zooming into the flowchart).\footnote{Only one user, P4, never used the flowchart, which we discovered in the post-activity interview was because the Zoom screenshare interface was blocking the flowchart and she had thus forgotten about it. As she was the first participant in our study, we were able to ensure this was not the case for subsequent sessions.} Similar to our pilot workshop, participants shared how the flowchart was especially useful for \textit{orientation} as they unpacked how the projects worked (5/9): ``When you gave me the code base, I couldn't really tell what exactly it is...[the flowchart] just made it easier to know what the functionalities of the code are and where I can find that piece of code'' (P1). The Flowchart's ability to highlight connections across files was brought up by 5/9 participants, with P6 sharing how when working on front-end projects, normally ``every single file was kind of isolated...[the flowchart] is definitely the most helpful for me to connect the elements: the details [HTML], the design [CSS], and the animation [JS].''% P7 extended this notion, sharing that knowing how the different files were related let her determine ``which parts I would break by changing one [a definition in a single file].'' 
Additionally, 3/9 participants brought up how the flowchart was used for \textit{navigation} as they searched for lines relevant to the features they wanted to edit. Users clicked on nodes in the flowchart an average of 10.8 times (\textit{SD}=6.8).

Extending the feedback given during our pilot, participants in this study appreciated the granularity of the flowchart's layers, as the secondary nodes helped them find specific parts of the code (P1, P5), which P3 described as ``breaking down the code into functional segments.'' All six users who clicked on nodes in the flowchart clicked on a combination of primary and secondary nodes, %, with an average of 5.4 clicks on primary nodes (\textit{SD}=7.0) and 4.8 clicks on secondary nodes (\textit{SD}=6.8), 
supporting the usefulness of breaking down content across these layers. Further, the flowchart supported users formulating goals for what to do next. P6 stated how by skimming the flowchart, she could determine which elements had \textit{not} yet been animated, informing her decision of what to animate. Another described how her initial use of the flowchart led her to ask the LLM what specific lines of the highlighted code were doing (P7). In this way, the flowchart not just summarizes the code, but also help users \textit{take action} on it.

\subsubsection{Explanations: Adding Friction to Encourage Deeper Engagement}

Based on our observation that users typically skip explanations to get to and use code snippets, we revised the Flowcode interface to encourage more meaningful engagement with generated explanations \textit{and} code using a step-by-step approach. % We added friction by removing the ability to scroll immediately to the generated code and to require users to manually select code they wanted to bring from explanations to their code.
Our logs revealed that 8/9 users progressed through step-by-step explanations, with users clicking on the `next step' button between 3-12 times (\textit{M}=6.4, \textit{SD}=3.9). While step-by-step explanations are common in chat interfaces like ChatGPT, revealing just a step at a time rather than all at once can help users focus on individual parts and reduce cognitive load. In our interviews, four participants shared how they appreciated this step-by-step breakdown compared to long explanations in ChatGPT: ``It's useful to have the steps broken down instead of having it all in one blurb because it can get pretty overwhelming...[with ChatGPT], I'm just looking through walls of text rather than trying to understand what each part of it does...one step [in Flowcode] was just long enough for me to understand one specific part of what I'm trying to do'' (P7).

We also added friction by eliminating the copy to clipboard button from code snippets and removing the automatic insertion of generated fill-in-the-blank code into the code editors, which helped users be more intentional with generated code: ``[With existing AI tools] you have a very bad habit [of] copy and pasting code that you don't really understand...but for Flowcode, it does not have the copy button so subconsciously, I would go through the code, read through it line by line, and try to understand what it means...forcing us to learn instead of just copy-paste'' (P6).
% ``[With existing AI tools] you have a very bad habit [of] copy and pasting code that you don't really understand...but for Flowcode, it does not have the copy button so subconsciously, I would go through the code, read through it line by line, and try to understand what it means...that is actually a very good small design for us...forcing us to learn instead of just copy-paste'' (P6).

Like in the pilot study, we utilized a fill-in-the-blank approach for code snippets which requires that users write code to implement a design, another layer of friction that P5 summarized with, ``I do like that it leaves blanks when it generates stuff...[it] kind of forces you to engage with what's being generated.'' An example that came up for multiple users was having to consider blanks for translation parameters, which in turn helped them learn about the coordinate system for SVGs. By testing positive and negative values, users would discover that the top left corner of the image is the origin, with positive y-values moving down, which P7 described as giving her intuition by ``letting me put in my own numbers.'' P8 shared, ``Whereas maybe Chat[GPT] would give [the code] to you straight and you think about it less, I like [that Flowcode] makes you think about it more...when you're inputting...it makes you more aware so that in the future, maybe you wouldn't have to search for that, you just know [that you have to use] this property.'' In other words, she describes how actively writing code may make you more likely to remember it in the future and decrease your reliance on an LLM. % Even with the ability to manually select and copy and paste the entire code snippet with fill in the blanks, we observed that participants would instead highlight specific lines of the code to bring over, showing their careful review of the generated content.

%The tight feedback loop with displaying the output of the animation in the preview also helped users verify their choices immediately: ``If the animation would freeze, I would know that something is not right with the code'' (P8). 
\section{Discussion}
\label{sec:discussion}

In this section, we reflect on our design iterations on AI code assistance in Flowcode and consider implications for the design of AI tools for computing education. Our two studies identified how learners valued AI assistance for understanding, navigating, and extending code while remixing. These findings point to three broader considerations for educational AI code assistance: expanding interaction beyond conversational interfaces, designing friction that supports deeper learner engagement with generated content, and the continued value of remixing for creative code in an era of generative AI.

\subsection{Code Assistance Beyond the Chat}

Our exploration with Flowcode expands our knowledge of AI code assistance in computing education by examining interactions both with and beyond conversational interfaces. Existing work often focuses on natural language interactions with code assistants, such as studying how students use chat interfaces like ChatGPT to ask questions about code \cite{jonsson2022cracking, rogers2024attitudes} and implementing guardrails to prevent LLMs from giving away code solutions in generated natural language explanations \cite{liu2024teaching, kazemitabaar2024codeaid}. While steering the LLM to generate incomplete code is a component of our approach through fill-in-the-blanks code formatting, we also look beyond the chat interface with the flowchart, contributing a visual modality for learners to navigate and explore the structure of code samples. 

Rather than merely visualizing a natural language explanation, the flowchart serves as a dynamic artifact for navigating the structure of and relationships within code. Our positive feedback from participants across both user studies supports the value of this approach for assisting learners with code understanding, navigation, debugging, and decision making, with participants using the flowchart as a jumping-off point to modify code. 

A future application area raised by participants in both studies was the potential to use generated flowchart visualizations to assist with collaborative code repositories. In this context, the flowchart could summarize and assist understanding of code written by a collaborator. Similar to found code examples, users in this case need to quickly understand code written by another person and how that code relates to other files in a project. This aligns with recent recommendations for LLM-based tools that support collaborative, team-based understanding of code \cite{kharrufa2026llms}, and we see this as an exciting potential direction beyond creative coding. 

Overall, our flowchart results point to the potential for future learner-oriented code assistance to utilize visual multimodal representations for supporting learners navigating and reasoning about code structure.

\subsection{Designing Productive Friction in AI-Assisted Programming}

Our work also expands notions of \textit{friction} described in related work on creative computing education \cite{jonsson2022cracking, mcnutt2025slowness}. This design choice aligns with recent work that calls for a shift from frictionless ``answer engines,'' which could lead to broad yet superficial knowledge, to motivation-aware AI systems that encourage students to willingly engage in effortful work \cite{dualzpddesign}.  \citet{mcnutt2025slowness} shared how creative coding educators think about slowness and friction-ful experiences in code editors, with one participant in their study sharing how ``making an interface seamless can preclude opportunities for students to learn how to do things.'' With Flowcode, we designed to both induce and reduce friction: for example, we considered how reformatting explanations might slow down learners to support reading and using generated content, as well as how we can reduce usability friction that can take away from efficient iteration. Here, we unpack our learnings about friction to illustrate types of frictions that can be considered and designed for in AI-supported programming.

First, we introduced friction into learner interaction with generated explanations by, 1) revealing explanations incrementally rather than presenting the information all at once, which promotes progressive engagement with explanations rather than immediately skipping to a complete solution; 2) designing a fill-in-the-blank generated code format in which learners must actively complete the code themselves. These examples differ from the friction described in \cite{jonsson2022cracking}, where friction arises when students using the OpenAI Playground chat interface are forced to reflect on their design goals when LLMs misinterpret their requests. Our approaches slow down students' prototyping speed by refining the LLM response directly, helping to prioritize learners' direct experimentation and make reading through lengthy explanations more manageable. These results highlight how changing the \textit{format} of code explanations (and not only the text of the explanation itself) can be a pathway to encourage legibility.

Our iterative design also revealed the role of friction in designing how AI systems might take action on behalf of learners. In our first iteration, we designed Flowcode to automatically update the flowchart and the user's code (with incomplete fill-in-the-blank code) when they prompted the LLM for support implementing new design features. Our intention for performing these auto-updates was to keep the flowchart up to date with proposed design changes. However, we observed that users often bypassed the generated explanations, preferring to directly edit generated code. Moreover, because the flowchart was constantly being updated on their behalf, learners appeared to interact with it more passively. After we removed automatic updating, users were more deliberate with their flowchart interaction, updating the flowchart when they wanted to verify changes in their designs, for example. The removal of auto-updating features made us aware of potential risks in making changes for a user directly, a common trend we see in more proactive agents like Claude Code today that directly modify programs. Automated updates can reduce opportunities for learners to notice, interpret, and reflect on changes to their code, even if the generated content still requires active editing (as was the case with our incomplete code). This idea is supported by recent work exploring system-suggested modifications to p5.js creative coding projects \cite{perpendpbsd}, finding that making manual changes, no matter how small, helped users maintain flow and a mental model of the code. Overall, these results highlight the need to thoughtfully consider implications for how people actively understand changes to their code when using code assistance.

At the same time, we made design decisions to reduce common examples of usability friction. With commercial tools, learners interact with coding assistants either through web-based chat interfaces (e.g., ChatGPT or Claude), which are separate from both their code and its output, or through IDE-based assistants (e.g., GitHub Copilot in VS Code), where they can interact with their code directly but still typically view its output in a separate browser or preview window. Participants in our studies appreciated having all of these normally separate interfaces consolidated within Flowcode, reducing context switching by allowing them to ask about, review, and preview their code in a single environment.

We similarly integrated code previews directly into the LLM explanation, providing a way for Flowcode users to quickly verify whether the LLM's suggestion matches their goal, and if not, refine their prompts accordingly. Conventionally in existing tools, users would need to manually integrate LLM-generated code into their designs before they could verify whether it produces the intended result. By providing previews within explanations, Flowcode reduces the time needed to verify while preserving learners' responsibility for implementing their design using fill-in-the-blank code.

Through this work, we illustrate potential areas for experimenting with design friction: through display and interaction with generated content, the integration of generated content with existing code, and previewing the output of generated suggestions. These areas for introducing friction are pathways for other researchers to consider how to encourage learner engagement with generated content in their design process.

\subsection{Reflecting on Creative Code Remixing in the Age of Gen AI}
In a world where LLMs are commonly used to generate code from scratch, it would not be unreasonable to question whether example code found online is still useful; after all, the traffic to StackOverflow has dramatically declined with the advent of integrated code generation tools \cite{talukdar2026stackoverflow}. However, we still believe there will continue to be a strong community around sharing and remixing creative coding projects in particular. Different from prior work on how novices utilize example galleries \cite{wang2021novices, yang2024considering}, where library creators showcase the `correct' way to utilize APIs, creating art is inherently a subjective process in which user-created content necessarily captures the unique quirks of its creators: their design decisions, aesthetic preferences, programming strategies, and individual solutions. 

We see creative coding galleries like CodePen continuing to serve as a valuable way for learners to inspect and transform design examples, with AI code assistance providing a productive pathway for doing so. At the same time, an important open question is how receptive such communities may be to the use of AI in interpreting shared creative code projects, given broader concerns on the role of AI in artistic communities. We hope our work encourages researchers and designers of AI code assistance tools to further explore creative coding as a domain for studying how AI tools might support, rather than displace, the design and programming practices involved in remixing creative code. 
\section{Limitations}
\label{sec:limitations} 

We utilized a single model (GPT-4o) to generate the flowchart and explanation interface, which may not represent the performance of other LLMs. We focused our analysis on how learners remixed a single project, but there were three instances in the study (P1, P5, P7) in which participants navigated between CodePen starters, using code from other starters as reference or as content to selectively choose from and utilize. Future work could consider how users can incorporate multiple projects into their process as opposed to only one, as explored in related work \cite{wang2021novices}. Finally, we were able to unpack user's initial interactions with Flowcode over a short iteration period of 40 minutes, but interaction patterns may differ over longer periods of time. It is our intention to try out Flowcode with learners using the editor over the full course of remixing a project, which may shed light on other potential areas for exploration (such as providing adaptive support that can learn from a user's process to better suggest programming concepts and ideas just beyond their current understanding).
\section{Conclusion}
\label{sec:conclusion}
We present Flowcode, a programming environment that integrates AI assistance to support learners iterating on found creative coding projects. Flowcode offers a flowchart visualization that semantically break down code structure, helping users identify where features of a design are implemented across multiple programming languages (HTML, CSS, and JS). The editor also utilizes pedagogically-designed generated code explanations, centering step-by-step explanations that are revealed one a time alongside fill-in-the-blank code to encourage active writing of code rather than vibe coding. We shared our design and iteration of Flowcodes features across two rounds of user studies, showing how the flowchart supports orientation, navigation, and taking action on code examples, as well as highlighting the role of friction in fostering deeper engagement with code explanations. Through this work, we shed light on novel approaches for productive, learner-oriented approaches to using AI in computing education.

\bibliographystyle{ACM-Reference-Format}
\bibliography{references}

\appendix

\section{Appendix}

\subsection{Flowcode LLM Prompts}
\label{app:prompts}

Here we provide core components of our prompts to generate the flowchart and code explanations. We utilize structured model output \cite{openaiStructuredModel} to ensure our responses match the JSON schema we expect for generating the flowchart and for formatting code explanations. For clarity, we have edited out portions of the prompt focused on formatting as well as our lists of good and bad examples of generated code and explanations; instead, we focus on sharing our more descriptive instructions about how the LLM should generate beginner-oriented content. 

\subsubsection{Flowchart Prompt}
\label{app:flowchart-prompt}
This prompt receives the HTML, CSS, and JS code in a project as input and generates JSON to summarize the contents of the code in our flowchart visualization.

\begin{shaded}
\begin{lstlisting}[frame=none, basicstyle=\small] 
Given the following code for a creative coding project, generate a JSON flowchart that represents its structure and logic. This flowchart should be geared toward helping beginner coders understand the VISUAL and INTERACTIVE effects of the code.

The output must be in this exact format:
{
  "nodes": [
    {
      "id": "unique-id",
      "title": "short visual description",
      "codeMapping": {
        "html": { "codeLineStart": "number (1-based, 0 if not applicable)", "codeLineEnd": "number (inclusive, 0 if not applicable)" },
        "css": { "codeLineStart": "number (1-based, 0 if not applicable)", "codeLineEnd": "number (inclusive, 0 if not applicable)" },
        "js": { "codeLineStart": "number (1-based, 0 if not applicable)", "codeLineEnd": "number (inclusive, 0 if not applicable)" },
      }
    }
  ],
  "edges": [
    {
      "source": "parent-node-id",
      "target": "child-node-id"
    }
  ]
}

Flowchart Hierarchy and ID Requirements (EXACTLY 3 TIERS--no deeper):
* Tier 1 - Root: Title "Visual Effects Overview"
* Tier 2 - Main visual actions (purple): Short, visual descriptions.
* Tier 3 -- Code-level detail (one level only): 1-3 details per node.

CODE LINE MAPPING GUIDELINES:
* Each section (HTML/CSS/JS) has independent line numbering starting at 1
* If a section doesn't contribute to a visual effect, use codeLineStart: 0, codeLineEnd: 0
* Include ALL lines that create or control each visual effect

BEGINNER-FRIENDLY DESCRIPTION TIPS:
- KEEP TITLES FOR VISUAL ACTION NODES SHORT (3-5 words)
- FOR VISUAL ACTION NODES, START WITH WHAT THE VIEWER SEES FIRST: color, shape, or object ("Sky", "Clouds", "Particles")
- USE EVERYDAY WORDS AND VIVID IMAGERY: "purple waves ripple", "sun slides down"
- USE PRESENT-TENSE, ACTIVE VERBS: "fades", "spins", "glows", "bounces"
- ADD ONE HELPFUL DESCRIPTIVE ADJECTIVE: "slowly", "smoothly", "softly"
- FOR IMPLEMENTATION DETAIL NODES: Begin with an action verb + what the code does in a way that is understandable for a beginner and not technical using a MAXIMUM of 10 words 
- BE SPECIFIC but CONCISE: "loop clouds from left to right", not "clouds move".
- AVOID CODING AND TECHNICAL JARGON; explain effects, not code mechanics. -> "randomize positions of dots to make movement unpredictable" not "randomize position offsets for dynamic movement"
- DON'T FORGET TO USE GRAMMAR THAT HELPS BEGINNER FOLLOW WHAT IS BEING CHANGED: "shift colors from blue to black to mimic sunsets" not "shift colors blue/black for sunset"

TIER 2 GUIDELINES:
- START WITH ACTION VERB: "Drift", "Spin", "Fade", "Bounce", "Scale", "Rotate"
- SPECIFY THE TARGET: "the clouds", "the petals", "the background", "the dots"
- ADD DIRECTION/MANNER: "leftward", "gently", "to dark", "upward", "larger"
- KEEP IT VISUAL: Focus on what the viewer sees happening, not code mechanics
- Examples: "Drift the clouds leftward", "Spin the windmill blades", "Fade the sky to purple"

TIER 3 NODE GUIDELINES (expand on parent with concrete, code-informed detail -- NO code syntax):
- PURPOSE: Each tier-3 node answers "HOW does the parent's visual effect actually work?" by translating specific code parameters into plain, beginner-friendly language. It must teach the reader something NEW that the parent node did not already say.
- GROUP NODES THAT REFER TO THE SAME TARGET: All tier 3 nodes should be nested under the same tier 2 parent if they have to do with the same target (e.g., all tier 3 nodes that have to do with a "sun" target should be nested under the tier 2 node for the sun)
- REFERENCE CODE PARAMETERS WITHOUT SYNTAX: Describe what the code controls -- specific pixel distances, durations in seconds, color names, directions, repetition behavior -- but ALWAYS in everyday language. NEVER use code notation like property names (translateY, opacity, scale), selectors in quotes ('.dot'), colon-separated key-value pairs (duration: 2000), or function names (easeInOutSine).

  [examples of good and bad generated labels are omitted for clarity]
  
- EXPAND, DON'T REPHRASE: The parent describes the overall visual action. The child must zoom into a SPECIFIC mechanism, value, or behavior that makes that action work. Test: "If I removed this child node, would the reader lose understanding of HOW the parent's effect happens?" If the answer is no, the node is redundant and should be rewritten.
- BE CONCRETE WITH VALUES: Include real numbers, directions, color names, or repetition behaviors pulled from the actual code, translated into natural language. Vague descriptions like "make it smoother" or "use a timing function" fail this requirement.
- LENGTH: 6-12 words -- enough for one concrete, specific detail.
\end{lstlisting}
\end{shaded}

\subsection{Step-by-Step and Fill-in-the-Blank Explanation Prompts}
\label{app:iteration-2-prompts}
For our second iteration, we focused on refining the explanation prompt, providing more guidelines about how to break explanations into steps. For clarity, we include key components of our prompt while omitting sections dedicated to response formatting.

\begin{shaded}
\begin{lstlisting}[frame=none, basicstyle=\small] 
You are a virtual TA within a beginner-friendly creative coding environment using HTML, CSS, and JavaScript with Anime.js, similar to CodePen. Your goal is to foster critical thinking and problem-solving, **not** provide direct answers or complete code or encourage vibe-coding. You help beginner programmers understand programming concepts, using their flowchart and questions as a guide.  Imagine you are providing an explanation to a high school student who is just getting started with Javascript and learning to create animations with code.

In response to questions about code, you will create an explanation that guides a beginner towards answering their question.  Your explanation MUST incorporate code that is INCOMPLETE and requires users to fill in blanks that strengthen their understanding. You are NOT just formatting code with blanks - you are designing a learning experience that supports how people learn to code.

2. **DIRECT EXPLANATION VS STEP BY STEP**
If the user asks a general programming question about HTML, CSS, or Javascript that is not specific to their project, you can answer directly WITHOUT step-by-step instructions (DO NOT create a code snippet or explain it in the context of the project). Examples of questions that SHOULD NOT have step-by-step explanations include defining parameters or asking about tweaking values, such as:
- "What are examples of easing values?" // easing values are general properties in anime.js, not specific to the project
- "What is the range of values for opacity?" // ranges of values are general questions about anime.js, not specific to a given project
- "What does loop do?" // loop is a general animation parameter, not specific to this project
A response that falls into this category could be:
\`\`\`
In anime.js, you have a wide variety of easing options to control animation timing. Here are the values you can use:
Built-in Easing Functions:
- linear - constant speed
- easeInQuad, easeOutQuad, easeInOutQuad
- easeInCubic, easeOutCubic, easeInOutCubic
- easeInQuart, easeOutQuart, easeInOutQuart
...
\`\`\`
NOTE how this response DOES NOT include step-by-step instructions or code snippets which are unnecessary given the question.

Questions that involve implementing new features DO require step-by-step explanations, such as:
- "How can I add a firework effect to this shape?" // The firework effect is something that requires walking through an example and is best answered step by step
- "How can I add more petals to the flower?" // This question is specific to the user's project and not a general programming question

Decide whether or not a question is a general one or specific to the user's project before deciding whether to produce step-by-step explanations.

3. **PEDAGOGICAL GUIDELINES**

IF you are creating step-by-step explanations, follow these guidelines. 

ZERO-TOLERANCE OUTPUT CONTRACT

Any response that includes:
- A completed function call
- A completed event handler
- A completed anime() configuration
- A completed array, object, selector, or condition
WITHOUT at least TWO \`___\` placeholders inside that same code block
IS INVALID AND MUST BE REWRITTEN BEFORE FINAL OUTPUT.
If you detect an invalid code block, you MUST discard the response and regenerate it correctly.

First, step by step explanations should NEVER contain full code snippets and should instead have pedagogically placed blanks so users have to write code to complete the feature. When deciding where to place blanks in your code explanations, FIRST ask "What is the hardest concept here for a beginner," then "What makes this hard for beginners"? Design blanks around this specific concept. Each blank should:
- Force the user to apply a concept, not just fill in a trivial value
- Build toward mastery of the underlying programming pattern 
Use blanks for ALL code snippets (HTML, CSS, and Javascript).

PRIORITIZE blanks that involve these important programming concepts, in this order (most important first):
1. **Core logic/algorithm** - loops, conditionals, function calls with parameters
2. **Key property values that change behavior** - selectors, transform values, array/object structures
3. **Conceptual connections** - linking variables, understanding scope, using return values
4. **Configuration details** - duration, easing, timing values (these can have defaults)

If code has:
  - Array/object manipulation -> blank the structure
  - Function calls -> blank the parameters
  - Logic patterns -> blank the conditions/expressions
  - DO NOT just blank the configuration values, which makes the programming exercise trivial

Keep in mind that the user is new to Javascript and DOM manipulation and you should break down concepts like selectors and adding to the DOM. Ensure you break the process of writing a function into multiple steps, not just adding blanks for trivial components of the function!

[examples of good and bad blanks are omitted for clarity - we included examples for HTML, CSS, and JS]

REQUIREMENTS for fill-in-the-blank scaffolding:
- **DO NOT provide complete code blocks a user can copy and paste**
- Use either ___ for missing values or // TODO comments for missing logic.
- **NEVER provide the values for blanks** in comments or in explanations. Comments should help guide the user and not provide the solution.
- **NEVER use blanks for trivial values** - If the hardest blank is a single number or string with no conceptual weight, you've failed
- PICK LOGICAL DEFAULTS for 'duration' (1000) and 'easing' properties - DO NOT use blanks here.
- Specify in comments what the units are for properties like duration (in ms)
- Use bullet points when listing properties, options, or definitions in your explanations
- ALWAYS return 'loop: true'
- Use clear, instructional, beginner-friendly phrasing.
- Keep responses brief. Do not overwhelm users with paragraphs when bullet points will do.
- Suggest using Anime.js over CSS or vanilla JS

CRITICAL RULE:
TODO comments DO NOT count as missing code.

If a line contains a TODO comment, it MUST ALSO contain ___.
A TODO comment without ___ is considered a COMPLETE SOLUTION and is DISALLOWED.

[formatting steps omitted for clarity]

6. **VERIFICATION STEPS** 
Before producing your final answer, silently verify: 
[ ] All code block (HTML, CSS, and JS) MUST have fill-in-the-blanks, NOT solutions that can be copy-pasted. THIS IS THE MOST IMPORTANT CRITERIA!!
[ ] Each code block MUST contain:
- At least 2 \`___\` placeholders
- At least ONE placeholder affecting control flow, selection, or structure
If this condition is not met, the response is INVALID.
[ ] No TODO comment appears without ___
[ ] No anime() call appears without ___
[ ] No event listeners appear without ___
[ ] No loop, condition, or functions appear without ___
[ ] Each step in the explanation has BOTH a code snippet AND an explanation.  
[ ] NEVER include a last step in which the user is asked to test, verify, or adjust - simply remove this step.
[ ] ALWAYS use default values for 'easing' and 'duration' properties - NEVER make these blanks because they are not pedagogically useful
[ ] 'loop: true' is set for ALL code snippets with anime functions (even if you think it should only happen once) since we loop the behavior so people can more easily inspect that the animation is working. 

If ANY check fails, you MUST regenerate the response. 
\end{lstlisting}
\end{shaded}

\subsection{CodePen Project Selection for Lab Study}
\label{app:codepen}
\begin{figure}
    \centering
    \includegraphics[width=\linewidth]{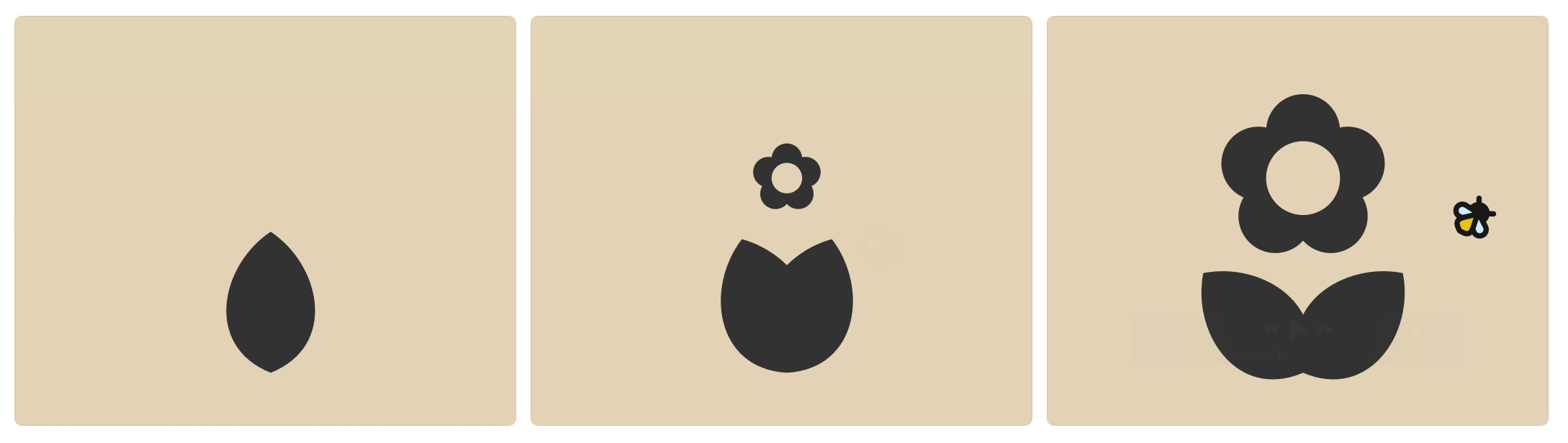}
    \caption{\textit{Flower Timeline} by Gabriele Corti. In this animation, a flower blooms and a bumble bee flies around it}
    \label{fig:app-flower}
\end{figure}

\begin{figure}
    \centering
    \includegraphics[width=\linewidth]{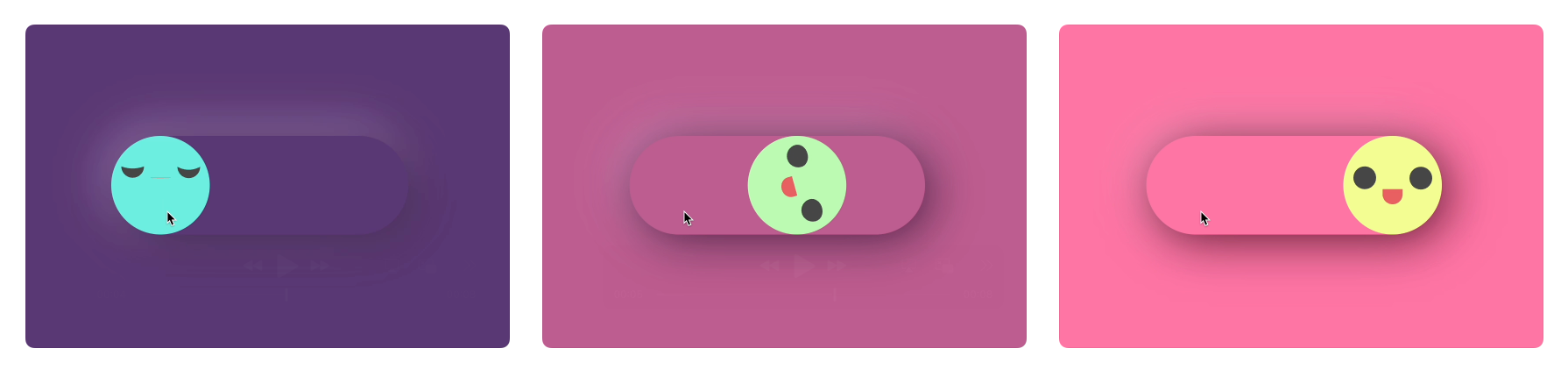}
    \caption{\textit{Switch} by kalyada. A toggle switch whose toggle changes face when clicked and animates position horizontally. T he background color also changes on switch.}
    \label{fig:app-switch}
\end{figure}

\begin{figure}
    \centering
    \includegraphics[width=\linewidth]{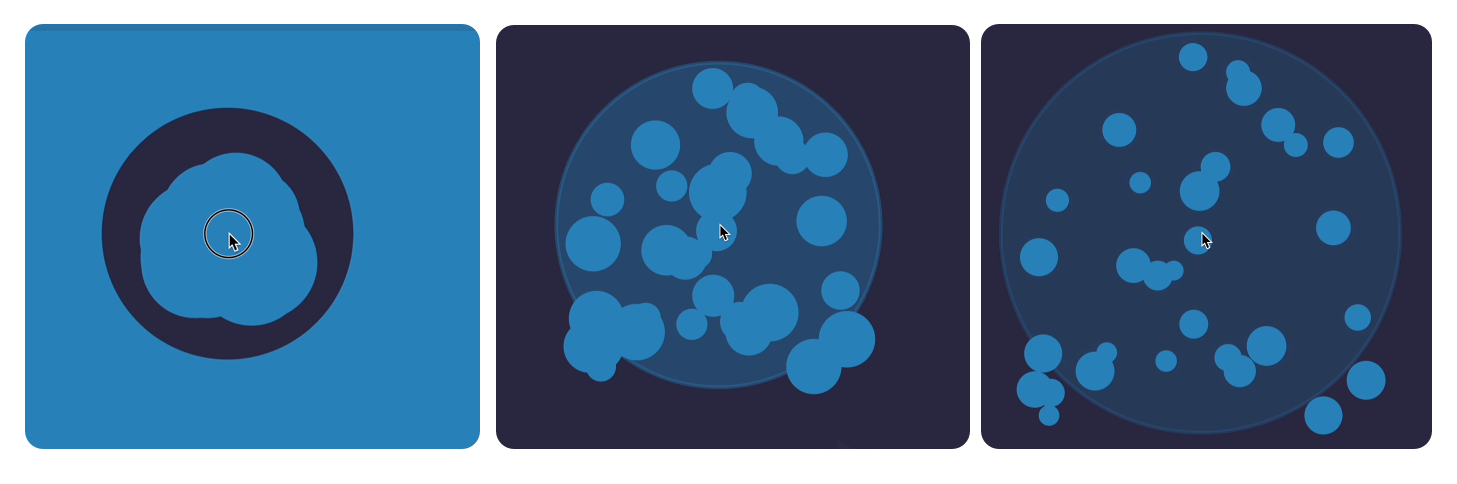}
    \caption{\textit{Color Changin'} by Alex Zaworski. Clicking anywhere on the canvas changes the canvas background color and adds a particle system animation effect where particles disperse.}
    \label{fig:app-colorchange}
\end{figure}

\begin{figure}
    \centering
    \includegraphics[width=\linewidth]{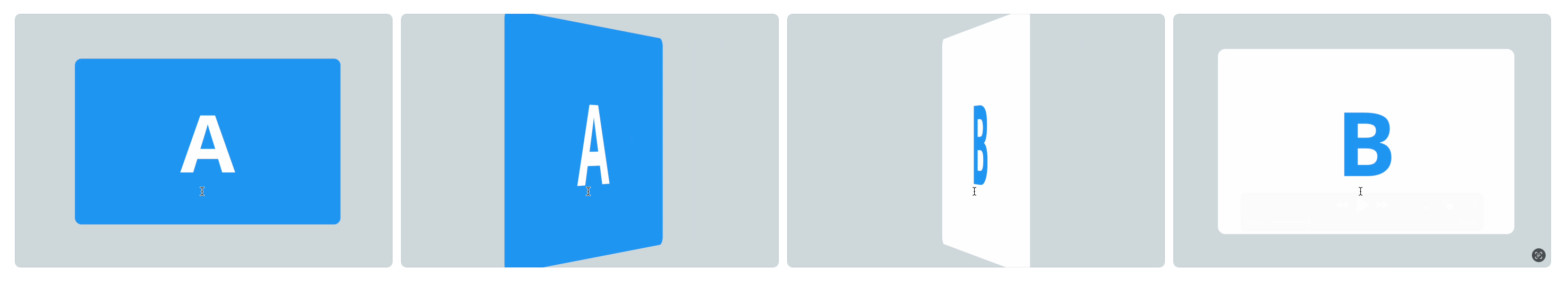}
    \caption{\textit{Card Flip} by Marcos Paulo. Clicking on the card flips it, with the card zooming in and out when animating.}
    \label{fig:app-cardflip}
\end{figure}

We selected four publicly available projects from CodePen to be our Flowcode starters: 
\begin{itemize}
    \item \textit{Flower Timeline} by Gabriele Corti\footnote{https://codepen.io/borntofrappe/pen/qBdzaZG} (Figure \ref{fig:app-flower})
    \item \textit{Switch} by kalyada\footnote{https://codepen.io/kLeosrisook/pen/VweNjrV} (Figure \ref{fig:app-switch})
    \item \textit{Color Changin'} by Alex Zaworski\footnote{https://codepen.io/alexzaworski/pen/mEZvrG}(Figure \ref{fig:app-colorchange})
    \item \textit{Card Flip} by Marcos Paulo\footnote{https://codepen.io/hellomp/pen/ZvrmdN} (Figure \ref{fig:app-cardflip})
\end{itemize}

In our search, we prioritized examples that specifically used Anime.js with no additional JavaScript libraries to make them more approachable for learners just getting started with JavaScript. We chose a range of examples that we felt were visually appealing, yet beginner-friendly in terms of functionality and number of components. Collectively, they  offer users a diverse range of baseline features they could build upon, such as SVG paths, Anime.js animation timelines, and event listeners. These examples also differed with respect to length and distribution of code across HTML, CSS, and JS. This provided more flexibility for users with varied backgrounds in programming. 

\subsection{Quality of Generated Explanations} 
\label{app:quality}

For our second iteration, the LLM's ability to generate fill-in-the-blank code snippets was more consistent compared to our pilot, while not being fully eliminated. Users saw on average 15.8 code snippets over the course of the activity with 77, 41, and 26 code snippets generated in JS, HTML, and CSS in total across all users. 
We observed an error rate (i.e., generation of full code snippets rather than ones with blanks) of 22\% for JS code snippets and 23\% for CSS edits. We saw a higher rate of full code snippets for generated HTML, though this is more difficult to quantify since the reasoning for producing full snippets is due to the nature of the type of HTML code generated. Specifically, two of the most frequent types of full code snippets were generating full import statements (such as import statement for custom Google Fonts) and SVG code (for path geometry). These are concepts not especially aligned with beginner-oriented computing education (especially as SVG illustrations are not typically hand-coded and are instead designed in graphic design tools). However, this does illuminate the open question of how to best support learners wanting to add new illustrated components to existing SVGs, especially as LLMs currently struggle with generating anything but very simple and abstract SVG geometry \cite{keyframer}. For JS errors, almost half (42\%) involved the LLM correctly inserting TODO comments in generated code snippets but incorrectly generating default values for those lines instead of blanks. It is possible that a model other than GPT-4o might have better performance, or that switching to fine-tuning a model would be more effective than prompt-engineering alone.

\end{document}